# A Two-Dimensional Propensity Score Matching Method for Longitudinal Quasi-Experimental Studies: A Focus on Travel Behavior and the Built Environment

**Haotian Zhong[1], Wei Li[1,3, *], Marlon G. Boarnet[2]**

[1] Department of Landscape Architecture and Urban Planning, College of Architecture, Texas A&M University
[2] Department of Urban Planning and Spatial Analysis, Sol Price School of Public Policy, University of Southern California
[3] Texas A&M Transportation Institute, Texas A&M University System
* Corresponding Author: wli@tamu.edu

October 31, 2019

**Abstract**
The lack of longitudinal studies of the relationship between the built environment and travel behavior has been widely discussed in the literature. This paper discusses how standard propensity score matching estimators can be extended to enable such studies by pairing observations across two dimensions: longitudinal and cross-sectional. Researchers mimic randomized controlled trials (RCTs) and match observations in both dimensions, to find synthetic control groups that are similar to the treatment group *and* to match subjects synthetically across before-treatment and after-treatment time periods. We call this a two-dimensional propensity score matching (2DPSM). This method demonstrates superior performance for estimating treatment effects based on Monte Carlo evidence. A near-term opportunity for such matching is identifying the impact of transportation infrastructure on travel behavior.

## 1. Introduction

Travel behavior and the built environment have direct and indirect implications for public health and climate change. Despite growing empirical evidence on the association between travel behavior and the built environment, this evidence is mostly based on cross-sectional studies that offer a limited causal understanding of the impact of the improvements to the built environment. Built environment research, as social sciences at large, faces multiple challenges for causal inferences. First, randomized controlled experiments (RCTs) have established as the gold standard for estimating causal effects, but they are often either infeasible or ethically problematic in planning research (Honey-Rosés and Stevens, 2017). Surely, we cannot randomize many policies and infrastructures; for example, we cannot randomize Zoning bylaws throughout a jurisdiction, or we cannot randomly assign households to poor or rich neighborhoods. Although we can conduct randomized experiments on some social programs, such as the Moving to Opportunity, those efforts usually are costly and require many years to know the outcomes, which may in turn provide less applicable implications for today's policies and practices.

Second, when true experiments are infeasible, researchers rely on natural experiments (NEs) to evaluate policy interventions, in which researchers have no control over who receives the interventions or expose to the improvements. One key evaluation challenge in NEs is the selection bias to the intervention, resulting in individuals or groups affected by the interventions that may be different from those unaffected in characteristics associated with the outcomes.[1]  For example, residents who live around a transit station can be different from those who live farther away due to location specific factors or be newly move in due to the preference of taking public transportation. This phenomenon is well acknowledged as residential self-selection bias in travel behavior research and more broadly known as intention-to-treat (ITT) effect in observational studies, leading to an estimate of the effect of group differences, rather than the effect of the intervention (Boarnet, 2011). Several methods have been used to address the selection bias, such as instrumental variables (IVs), regression-discontinuity (RD) design, synthetic control, and matching. When studying the behavioral impacts of transportation improvements on individuals, the strength of those methods is tempered by the lack of true panel data that track the same individuals over time.

While the major drawback of transportation panel data is that they are rarely available, the repeated cross-sectional (RCS) data are abundant, such as National Household Travel Survey (NHTS), American Community Survey (ACS), and travel surveys conducted by Metropolitan Planning Organizations (MPOs). Unfortunately, many RCS data are underutilized in developing causal understandings of travel behavior and the built environment. Yet, RCS data have been used in many studies for identifying patterns or controlling for confounding effects, but they have not been widely used in causal estimations due to the changes of the before-after and treated-control groups in response to interventions are potentially incomparable. Furthermore, RCS data have the potential to substantially advance the causal understandings of travel behavior and the built environment — if a quasi-experimental setting can be constructed. Others proposed different methods to construct pseudo-panel data from RCS data. For example, Deaton (1985) introduced the pseudo-panel data analysis to consumer demand systems; Dargay and Vythoulkas (1999) adapted this approach in transportation researches. These studies mainly constructed pseudo-panels based on age cohorts from multiple rounds of RCS data. Verbeek and Vella (2005) reviewed more similar studies and concluded that these studies employed essentially the same

---

[1] Studies may suffer reverse causality issues when relying on cross-sectional data for analysis. Of course, the omitted variable issue always exists in any study

model that is the first order autoregressive model with exogenous variables. Unfortunately, these estimators cannot be implemented under general conditions due to the unsatisfied instrumental variable conditions in practice, as indicated by Verbeek and Vella (2005).

With this perspective, we aim to develop an approach for constructing pseudo-panel data from the abundant RCS data. In this article, we present and evaluate an array of matching algorithms for two-dimensional matching, with a focus on travel behavior and the built environment. We design two-dimensional propensity score matching (2DPSM) as a tool to pair individuals between the treated and control groups at both longitudinal and cross-sectional dimensions based on their characteristics. 2DPSM is designed to address the two biases:

First, *selection bias*: the bias results from the estimated samples are nonrandomly assigned into treatment group and control group due to self-selection.

Second, *longitudinal incomparability*: the bias results from the estimated samples may be systematically different before and after the intervention due to no true panel data.

Therefore, 2DPSM constructs a longitudinal quasi-experimental setting by mimicking a random assignment at the cross-sectional dimension and pairing statistically identical individuals at over time. This approach cost-effectively enables researchers to establish causality between improvements in the built environment and behavioral outcomes by using widely available RCS survey data. For example, our 2DPSM method has the potential to longitudinally match respondents from two different travel surveys which take place a few years apart for the same region; the method can help construct a quasi-experimental setting based on respondents from the two surveys and assess the behavioral impact of a major transportation project (e.g., light rail transit and bus rapid transit) that becomes open between the two surveys.

The rest of the article is organized as follows. Section 2 presents our two-dimensional matching framework. Section 3 provides a general review of causal inference frameworks and methods. In Section 4, we apply our proposed framework to numerical experiments to illustrate the performance of our methods. Section 5 discusses the strengths and limitations of 2DPSM and provides step-by-step guidelines to researchers who are interested in its applications. Section 6 concludes.

## 2. Two-Dimensional Propensity Score Matching

### 2.1 Biases in Quantities of Interest

To illustrate the 2DPSM framework, we briefly describe a typical example of travel behavior and the built environment research. Consider a public transit system opening in the neighborhood as treatment, individuals within the catchment area of public transit stations as being treated, and ridership as the outcome. Policymakers and researchers in general are interested in the effects of such transit station openings on residents and local businesses. Of course, such an effect can be dynamic, heterogeneous, and nonlinear, depending on contexts and time.[2] Here, we focus on the average treatment effect, with a focus on individual behaviors. Therefore, in a typical setting, we have four groups of people related to opening of a new transit system: (1) before the opening and within the catchment area (before and potentially treated=BT); (2) before the opening and beyond the catchment area (before and control=BC); (3) after the opening and within the catchment area (after and treated=AT); and (4) after the opening of the transit system and beyond the catchment area (after and control=AC).

---

[2] We agree with Chatman (2014) that treatment effect can be heterogeneous across population groups, and residential self-selection can inform policy questions. We also argue that the treatment effect can be dynamic and nonlinear over time. Researchers should carefully design their research based on what specific questions they are asking. Here, we focus on the application in estimating the average treatment effect.

Suppose we have individuals i (i= 1 to n) and their associated travel behaviors at time t, taking the new public transit $Y_{it}(1)$ and not taking $Y_{it}(0)$. The treatment effect (TE) on an individual is defined as $Y_{it}(1) - Y_{it}(0)$. First, we assume that $Y_{it}$ is given by a linear factor model.

$$Y_{it} = \delta_{it}T_{it} + x'_{it}\beta + \lambda'_i f_t + \varepsilon_{it} \tag{1}$$

where the treatment indicator $T_{it}$ equals 1 if individual *i* has been exposed to the treatment prior to time t and equals 0 otherwise; $\delta_{it}$ is the heterogeneous treatment effect on individual i at time t. The average treatment effect (ATE) is defined to be E $[Y_{it}(1) - Y_{it}(0)]$, which is the average effect on all individuals. The "average treatment effect on the treated" (ATT) is defined to be E $[Y_i(1) - Y_i(0) \mid T_i=1]$, which is the average effect on individuals in the treatment group. In practice, the ATE and ATT are in-sample estimates, namely sample average treatment effect (SATE) and sample average treatment effect on the treated (SATT). Imbens (2004), Kurth et al. (2006), and Imai et al. (2008) have provided further discussion for the distinctions. We focus on ATT in this paper for simplicity, which is the average treatment effect overall treated individuals in the sample. Given our earlier discussion on ideal RCTs, the standard Difference-in-Differences estimator can be written as:

$$SATT = E[Y_{it1}(1) - Y_{it1}(0)] - E[Y_{it0}(1) - Y_{it0}(0)] \tag{2}$$

Time periods *t₁ and t₀* denote before and after treatment, respectively. In reality, one individual can only be in the control group or treatment group. Therefore, we can only observe one outcome for the individual *i*: *Y_i (1)* or *Y_i (0)*. In the treatment group, *Y_i (1)* is observed, and the *Y_i (0)* is unobservable. The *Y_i (0)* will be replaced by the outcome *Y_j (0)* of a matched subject *j* from the control group, where *selection bias* may arise. Moreover, as individuals are not tracked over time in RCS, compositions of the treatment group and control group are likely to be different after interventions, where *longitudinal incomparability* may arise.

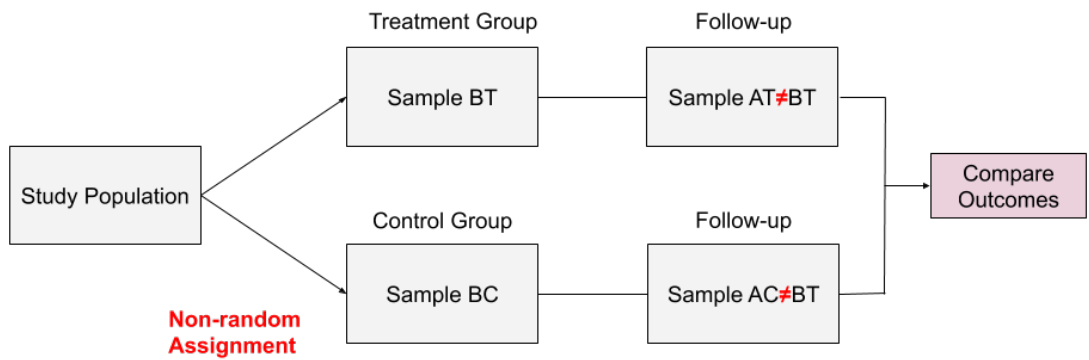

Figure 1 A typical setting of travel behavior and the built environment research

### 2.2 Two-Dimensional Propensity Score Matching Framework

Two-dimensional propensity score matching is proposed to mimic a quasi-experimental situation wherein the four groups of individuals with balanced covariates distribution (see Figure 1). The two main estimation bias sources are the selection bias that individuals with certain characteristics self-select to the treatment group and the longitudinal incomparability that individuals may not be the same groups of people after the intervention. At the cross-sectional dimension, we match individuals have similar propensity scores, which is the probability of being in the treatment group conditional on their characteristics. At the longitudinal dimension, we match individuals with identical characteristics. The matchings at both cross-sectional and longitudinal dimensions are expected to reduce estimation bias result from self-selection and incomparability.

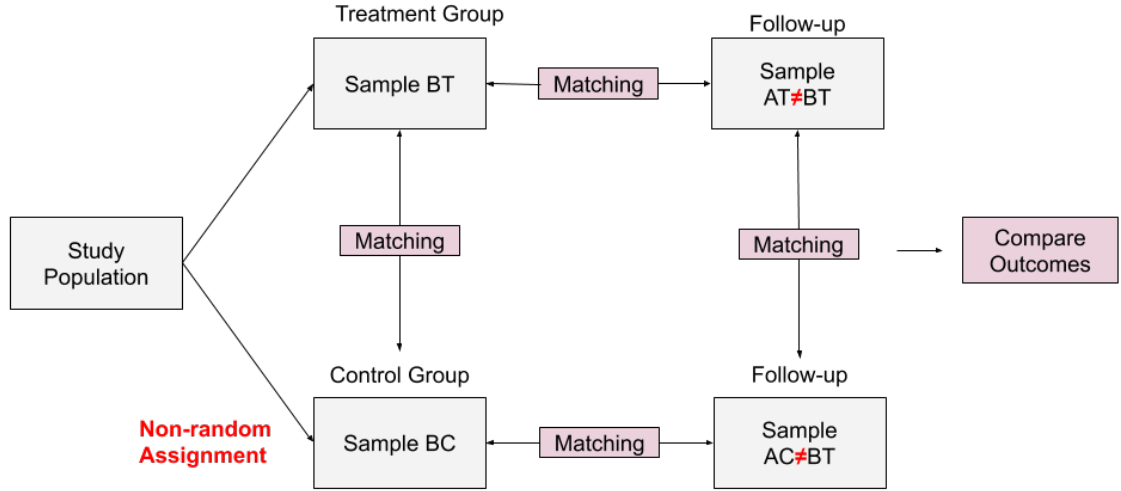

Figure 2 Proposed two-dimensional propensity score matching for causal inference

Matching plays the central role of summarizing all the relevant information and balancing the set of observed covariates across groups. After matching at both longitudinal and cross-sectional dimensions, Equation (2) can be rewritten as:

$$D_0 = E\big[Y_{it0matched}(1) - Y_{jt0matched}(0)\big|p(X_a^b),\ p(X_c^t)\big] \tag{3}$$

$$D_1 = E\big[Y_{it1matched}(1) - Y_{jt1matched}(0)\big|p(X_a^b),\ p(X_c^t)\big] \tag{4}$$

$$SATT = (D_1 - D_0)\ |\ X \tag{5}$$

$D_0$ and $D_1$ are the differences in means between treatment and control groups before and after treatment, respectively. The $p(X_a^b)$ and $p(X_c^t)$ denote matching at cross-sectional (b: before treatment, a: after treatment) and longitudinal dimensions (t: before and after treated groups, c: before and after control groups). When matching is properly applied at both dimensions, the before-after and treated-control groups are balanced with similar distributions of covariates. In theory, the difference between $D_1$ and $D_0$ is defined as SATT over the common support of X. The underlying assumption is the common support among the four groups must be met. In travel behavior researches, the subjects grouped by geographical areas and temporal periods have no dramatic differences in most cases. The application of this framework also requires the usual ignorability assumption, that is the selection or assignment of the before-after and treated-control groups only depend on the observed characteristics of individuals (Rubin, 1977).

**2.3 Matching Methods**

Matching at longitudinal and cross-sectional dimensions could involve several methods that can balance the distributions of covariates in treated and control groups, mainly optimal matching and greedy nearest neighbor matching (Guo and Fraser, 2014). Optimal matching pairs subjects to minimize the average within-pair differences in propensity score for the whole groups. In contrast, greedy nearest neighbor matching selects a treated subject and then pair it with an untreated subject with a propensity score closest to that of the treated subject. Also, greedy nearest neighbor matching (NNM) has several varieties by varying whether drop off subjects during the matching (i.e., with or without replacement), prune matched pairs beyond a chosen caliper (caliper matching), and whether match one subject to another, which is usually preferred, or to many others (one-to-one or many-to-one matching). Austin (2011a, 2014) has a detail introduction and comparison of each matching estimators. We develop and evaluate 2DPSM based on one-to-one nearest neighbor matching.

NNM is based on distance metrics, mainly propensity score matching (PSM) and Mahalanobis distance matching (MDM). PSM reduces the multi-dimensional vectors to a scalar propensity score. The propensity score is the probability of one subject receives the treatment conditional on observed baseline covariates, as defined by Rosenbaum and Rubin (1983). PSM pairs treated and untreated subjects with a similar value of propensity score (Rosenbaum and Rubin, 1983, 1985). The concept of Mahalanobis distance was proposed by Mahalanobis (1936). It calculates how many standard deviations away from a point to the mean of distributions. The MDM was invented even earlier than the propensity score matching by Cochran and Rubin (1973) and Rubin (1980). The MDM chooses the subject in the control group with the minimum number of standard deviations of Mahalanobis as the match for the treated subject.

**3. Causal Inference in Travel Behavior and the Built Environment Research**

The key idea is that a causal effect is a theoretical quantity and independent from any empirical methods that used to estimate it. This idea originates from many others, for instance, Rubin (1974) in statistics, Roy (1951) in econometrics, King et al. (1994) in social science, and Lewis (1973) in philosophy. The causal effect is inherently to compare the potential outcomes of individual $i$ being treated $Yi$ $(1)$ and untreated $Yi$ $(0)$.  Ho et al. (2007) provided an example of a fixed causal effect that is the causal effect at observation level, and a random causal effect is the average causal effect. In most applications, we do not seek to estimate fixed causal effects for each observation. Instead, we are interested in estimating the average effect over a particular subset of the population say the average increased walking trips of individuals live in new urbanism design neighborhoods comparing with individuals live in traditional suburb neighborhoods. Randomized controlled trials (RCTs) are considered the golden standard causal inference framework.[3] Three critical features guarantee valid causal inferences: (1) random selection of sample, (2) random assignment of treatment, (3) large sample size. The social experiment method is a similar method to classical randomized experiment and rarely conducted in travel behavior and the built environment research (e.g., Spears et al. (2016)). Although the social experiment method is most convincing in social science research, it almost always fails to meet the experimental conditions, for example, the sample selection might not be random, and the sample size is often small (Blundell and Dias, 2009).

---

[3] We would like to acknowledge the significance and contributions of other causal inference frameworks, including Granger causality, causal structure learning framework, causal graphical models, and nonlinear state-space methods. We refer to Heinze-Deml et al. (2018)  for a broader understanding of causal inference. Here, we develop our method under the Neyman-Rubin potential outcome framework (Holland, 1986; Rubin, 1974, 2005; Splawa-Neyman et al., 1923).

In observational research, natural experiments are widely accepted and implemented. The natural experiment approach simply compares the difference in means of pre-treatment and post-treatment outcomes with comparison groups. This method is often labeled as difference-in-differences (DID) estimator, which has been widely applied and improved by researchers(Abadie, 2005; Card, 1990; Card and Krueger, 1993; Garvey and Hanka, 1999; Heckman et al., 1997b; Imbens et al., 1997; Meyer, 1995; Meyer et al., 1995). DID estimators can control for temporal effects that bias the treatment effects and unobservable individual effects, relying on two strong assumptions: (1) treated and control groups would have a common trend in the absence of treatment, and (2) no systematic composition changes over time within each group.

Unfortunately, observations are not selected into the treatment group randomly in travel behavior and the built environment research. For example, individuals who live around transit stations are usually systemically different from people who live farther away. Two factors contribute to this issue. Firstly, individuals with a higher propensity to use the transit system would self-select to live near transit stations. The higher propensity is usually associated with lower earning, physical difficulty of driving, and active travel attitudes. Secondly, the route selection of transit systems is not random. Transit stations are usually placed in locations that have more demand for public transportation. As a consequence, the size and composition of the treated group may not be comparable over time, which could contaminate the true treatment effects. Most often, researchers have conducted quasi-experimental or quasi-longitudinal design or used advanced econometric techniques based on cross-sectional data to accommodate the limitations(McCormack and Shiell, 2011). For example, Cao and Schoner (2014) compared ridership of residents along the Hiawatha line in Minneapolis with residents who live in a similar corridor using propensity score matching. Handy et al. (2006) studied the relationship between the built environment and walking in a quasi-longitudinal framework. However, the direction of the causal relationship is still hard to be determined from cross-sectional data (Mokhtarian and Cao, 2008). The problem of reverse causality remains. Moreover, even if researchers can collect before and after data, the repeated observations in observational research are usually not from the same individuals.

Matching methods are widely used to replicate a randomized experiment as closely as possible, accounting for the systematical differences in treated and control groups (Stuart, 2010). In the extraordinary case where data are exactly matched, the causal effect of a treatment, policy intervention, or exposure can be estimated by comparing the means of treated and control groups. However, matching is usually used to balance the multivariate distributions of the treated and control groups and maximize the size of the matched data set. Therefore, matching has been considered as a pre-processing technique to reduce model dependence rather than a method of estimation, and any remaining imbalance need to be dealt by the best parametric procedures as used without matching (Cochran and Rubin, 1973; Ho et al., 2007; Rosenbaum and Rubin, 1985; Rubin, 1980). Matching has started to be utilized as an important tool in travel behavior research recently, especially those of the interest of policy evaluation. For example, researchers have utilized propensity score matching methods to recover various policy effects, such as transit ridership (Cao and Schoner, 2014), impacts of density and neighborhood type on travel behavior(Cao and Fan, 2012; Cao, 2010), connections among self-selection, residential location, and driving (Cao et al., 2010), and the impact walkability on housing market(Xu et al., 2018). Despite the significance of matching methods employed in previous studies, these applications either have unique research contexts or panel data (e.g., travel diary) that do not require two-dimensional matching.

## 4. Matching Protocol

In this section, we describe the matching protocol for 2DPSM. At the cross-sectional dimension, PSM is used for matching, as selection bias is essentially the unequal probability being assigned to treatment and control groups. At the longitudinal dimension, we use either PSM and MDM to minimize the differences in observed characteristics of individuals.

2DPSM Framework: A matching procedure that sequentially pair individuals in each before-after and treated-control group takes the following steps:

---

**Step 1. Start** the first round of matching for each of the group

  (a)  Start matching individuals in BT with individuals in AT
  (b)  Use matched individuals in BT to match with individuals in BC
  (c)  Use matched individuals in BC to match with individuals in AC
  (d)  Use matched individuals in AT to match with matched individuals in AC

  **End** of this step, collecting matched individuals in each group.

**Step 2. Start** a loop that runs R times:

  (a)  In round 1, repeat the procedure in step 1 with matched individuals
  (b)  Continue to round 2 if the numbers of matched individuals in each group do not equal. Otherwise, end the loop, collecting matched individuals in each group.
  (c)  In round $r \in \{2, \dots, R\}$, repeat the procedure in step 1 with matched individuals in the previous round.
  (d)  Continue the loop if the numbers of matched individuals in each group do not equal. Otherwise, end the loop, collecting matched individuals in each group.

  **End** of the loop.

**Step 3.** Check imbalance of the covariates.

  For cross-sectional dimension, check imbalance for BT : BC, AT : AC.
  For longitudinal dimension, check imbalance for BT :AT, BC : AC.

**Step 4.** Proceed to estimations

---

We specify both PSM and MDM with replacement. Thus, a treated subject is randomly selected and paired with an untreated subject whose propensity score is closest to the selected subject. Once an untreated subject matched with a treated subject, the untreated subject is dropped off the pool and will not be further considered in the matching process. To define the closeness, Rosenbaum and Rubin (1985) suggested a caliper value of 0.25 standard deviation of the propensity score, and Austin (2011b) suggested a caliper value of 0.2 standard deviation. Therefore, we test 2DPSM performance using the two different caliper values, adding no caliper, 0.2 standard deviation, and one standard deviation of the propensity score for the purpose of testing. The greedy matching is most suitable for the control group has a fairly larger number of subjects than the treatment group (Althauser and Rubin, 1970; Stuart, 2010). This situation holds in most of the public transit researches that only a small portion of the population lives within the catchment area of transit stations. Future researchers can apply caliper values based on the specific case, such as the nature of the research question, data quality, and data collection mechanism.

## 5. Evaluating Performance of 2DPSM

In the series of Monte Carlo simulations, we are interested in exploring the performance of the two schemes in estimating the FSATT, varying proportions of treated group, matching methods, and levels of imbalance before and after. Our aim for the Monte Carlo study is to answer following research questions: (1) Within each level of imbalance, how much imbalance can be reduced by each scheme? (2) Within each level of imbalance, which scheme performs the best, and what are the specifications in terms of matching methods and caliper values?

### 5.1 Data Generation Process

We generate data building upon Gu and Rosenbaum (1993) and King and Nielsen (2016). Four covariates (X1, X2, X3, X4) are drawn from a normal distribution with variances of 1 and covariances of 0.2 and 0.9. The level of balance between the potentially treated group and control group is high before treatment by setting the control group mean vector to (0, 0, 0, 0) and treatment group mean vector to (0.1, 0.1, 0.1, 0.1). The levels of balance between the treated group and other groups are very high, high, medium, low, and very low by setting the treated group mean vector to (0.1, 0.1, 0.1, 0.1), (0.3, 0.3, 0.3, 0.3), (0.5, 0.5, 0.5, 0.5), (1, 1, 1, 1), and (2, 2, 2, 2), respectively. For each level of balance, we generate 1,000,000 observations and draw 1,000 datasets by sampling 1,000 observations from the pool each time. Within the 1,000 observations in each dataset, the numbers of treated observations are 100, 300, and 500, respectively. The treatment effect is set to 0.6. The coefficients of the four covariates take the following values based on Austin (2011b): low effect log (1.25), medium effect log (1.5), high effect log (1.75), and very high effect log (2). The error term is drawn from normal distribution with a variance of 0.5 and a mean of 0. There are 15 settings by above specifications and are summarized in Table 1.

Table 1 Simulation settings by balance level and control-treated ratio

| Balance level | Control : Treated ratio = 9:1 | Control : Treated ratio = 7:3 | Control : Treated ratio = 1:1 |
|---|---|---|---|
| **Very low** | A0 | B0 | C0 |
| **Low** | A1 | B1 | C1 |
| **Medium** | A2 | B2 | C2 |
| **High** | A3 | B3 | C3 |
| **Very high** | A4 | B4 | C4 |

We also evaluate the performance of 2DPSM using the data generation process that defines the treatment assignment using a true propensity score model used in Setoguchi et al. (2008) and Lee et al. (2010) and set different means of vectors before and after. However, the data generation process cannot demonstrate the strengths of 2DPSM. We find that the DID alone can produce powerful estimations on the treatment effect using such a data generation process. The explanation could be that the means are different before and after but both in treatment and control groups, so the DID estimator can capture the changes in the fixed-effect.

### 5.2 Matching Scenarios

We conduct simulations for various specifications of 2DPSM. We report the following specifications for the purpose of illustrating combinations of matching techniques and range of caliper chosen. And another reason is that we would finally prune too many observations with a small caliper and as such no matched samples are generated, for example 2D-1 with a caliper of 0.2 standard deviation can drop all observations out during the matching process. We also compare 2DPSM with two other methods: (1) estimation without matching and (2) a common practice in the literature that only match individuals at cross-sectional dimension. We report each method and specifications of 2DPSM in Table 2.

Table 2 Matching Scenarios and Specifications of 2DPSM

| Matching Scheme | Notation | Description |
|---|---|---|
| No Matching | Naïve | No matching is implemented. |
| One-Dimensional Propensity Score Matching | 1D | PSM is implemented at cross-sectional dimension (caliper 0.2*SD), covariates are included for longitudinal comparison. |
| Two-Dimensional Propensity Score Matching | 2D-1 | PSM is implemented at cross-sectional dimension (caliper 1*SD), and MDM is implemented at longitudinal dimension. |
| | 2D-2 | PSM is implemented at both cross-sectional and longitudinal dimensions (caliper 0.2*SD). |
| | 2D-3 | PSM is implemented at both cross-sectional and longitudinal dimensions (caliper 1*SD). |

### 5.3 Checking Imbalance

We rely on the standard difference to measure the imbalance between treated and matched control (d'Agostino, 1998). The standardized difference is calculated as:

$$\delta = \frac{|\bar{x}_t - \bar{x}_c|}{\sqrt{\frac{s_t^2 + s_c^2}{2}}} \tag{6}$$

The conventional indicator of balance is a standardized difference $\delta \leq 10\%$ between treatment and control groups (Oakes and Johnson, 2006). We check the imbalance between the treated group after treatment and other groups. The imbalances of X1 and X2 without matching are range from 0.1 to 2 between the treated group after treatment and other groups, the other groups all have high balance initially. After matching, all schemes of 2DPSM reduce a significant degree of imbalance between the treated group after treatment and other groups. Method 1D fails to reduce the longitudinal imbalance that might be accommodated by regression analysis. Among all the specifications, 2D-2, which is with a smaller caliper, results in the most substantial reductions in the imbalance in most cases. See Appendix A for a detailed imbalance check.

### 5.4 Measuring Performance

We use four criteria to assess performance: matched sample size, estimated bias, estimated root mean square error (RMSE), and 95% confidence interval coverage of estimated treatment effect. The matched sample size is the final sample size after matching because a small matched sample size can inflate the variance of the causal effect. The estimated bias is based on the mean value of the estimated treatment effect of the 1,000,000 samples, which is calculated from the mean average treatment effect minors the treat treatment effect. We report the percentage of bias, which is the bias divided by the true treatment effect. The RMSE is estimated by the square root of the average squared differences between the estimated average treatment effect and the true treatment effect. The 95% confidence interval coverage of the estimated treatment effect calculates what percentage of estimated treatment effects fall within the 95% confidence interval of the true treatment effect. Figure 7 reports the performance measures of estimated treatment effects. Appendix B presents the numerical values of the performance measures.

**Matched sample size:** The matched sample sizes generally increase when the treatment prevalence increases, and decrease when the imbalance level increases. The matched sample sizes increased with the number of treated subjects, but small control groups can forbid the 2DPSM due to the difficulty of finding a match for treated subjects (see Table 3). For example, 2D-2 cannot run in the A2, A3, and A4 scenario. The higher the treatment prevalence results in a larger matched sample size since we aim to pair treated subjects and with just one control subject. At higher prevalence, the matched sample sizes decrease faster when the imbalance level increases than at lower prevalence because it is more likely to find a matched subject when the control group larger than the treatment group.

Table 3 2DPSM specifications applicability in each scenario

|  | A0 | A1 | A2 | A3 | A4 | B0 | B1 | B2 | B3 | B4 | C0 | C1 | C2 | C3 | C4 |
|---|---|---|---|---|---|---|---|---|---|---|---|---|---|---|---|
| **Naïve** | Y | Y | Y | Y | Y | Y | Y | Y | Y | Y | Y | Y | Y | Y | Y |
| **2D-1** | Y | NA | NA | NA | NA | NA | NA | NA | NA | NA | NA | NA | NA | NA | NA |
| **2D-2** | Y | Y | NA | NA | NA | Y | Y | Y | Y | NA | Y | Y | Y | Y | NA |
| **2D-3** | Y | Y | Y | Y | NA | Y | Y | Y | Y | NA | Y | Y | Y | Y | NA |
| **1D** | Y | Y | Y | Y | NA | Y | Y | Y | Y | NA | Y | Y | Y | Y | NA |

Note: Y denotes matching completed; NA denotes observation out of number during the matching process.

**Average treatment effect:** Mostly, 2DPSM improves the estimation of the true treatment effect. Recall the true treatment effect is 0.6. A horizontal dotted line is added to Estimated Treatment Effect panel denoting the magnitude of the true treatment effect. Results from Naive method are increasingly divergent from the true treatment effect as the imbalance level increases. Except for Naïve, other specifications have satisfactory estimated average treatment effects on treated. The estimated ATT is close to the true parameter when use 1D, largely because we included the covariates in the DID estimation. The estimated ATT are simply differences in means without controlling covariates in 2D-1, 2D-2, and 2D-3. 2D-1 outperforms 2D-2 in all scenarios that 2D-2 can be applied.

**Standard deviation and Bias ratio:** We report the standard deviation and bias ratio of the estimated treatment effects across the 1000 replications. In line with the trade-off between the bias and variance, we find that the larger bias ratios associate with the smaller standard deviations, comparing 2D-2 and 2D-3. In scenarios with low imbalance level, all methods yield little bias. In part because of DID does not require panel data if groups are similar over time. The Naïve estimation results in significantly biased results when imbalance levels are high.

**RMSE:** Again, 2DPSM performs better at the higher prevalence of treatment and lower imbalance level. Relative to the Naïve, the 2DPSM can reduce the RMSE by about 90%. 1D has the best performance. 2D-2 has a very similar performance to 2D-3.

**95% CI coverage:** In a majority of scenarios, 2DPSM achieves 95% CI coverage, about 95%. There is one pattern that bears noting. 2D-3 has a low 95% CI coverage in C1 and then return to medium and high coverage in C2 and C3. We can see that the 95% CI coverage increases as the matched sample size decrease. Since the 2D-3 chose one standard deviation of propensity score as the caliper, it is very likely to include less suitable matched subjects at 50% prevalence of treatment. At a higher level of imbalance, more irrelevant information is excluded, resulting in a smaller matched sample size as well as better estimation. This pattern shows in all the above measures

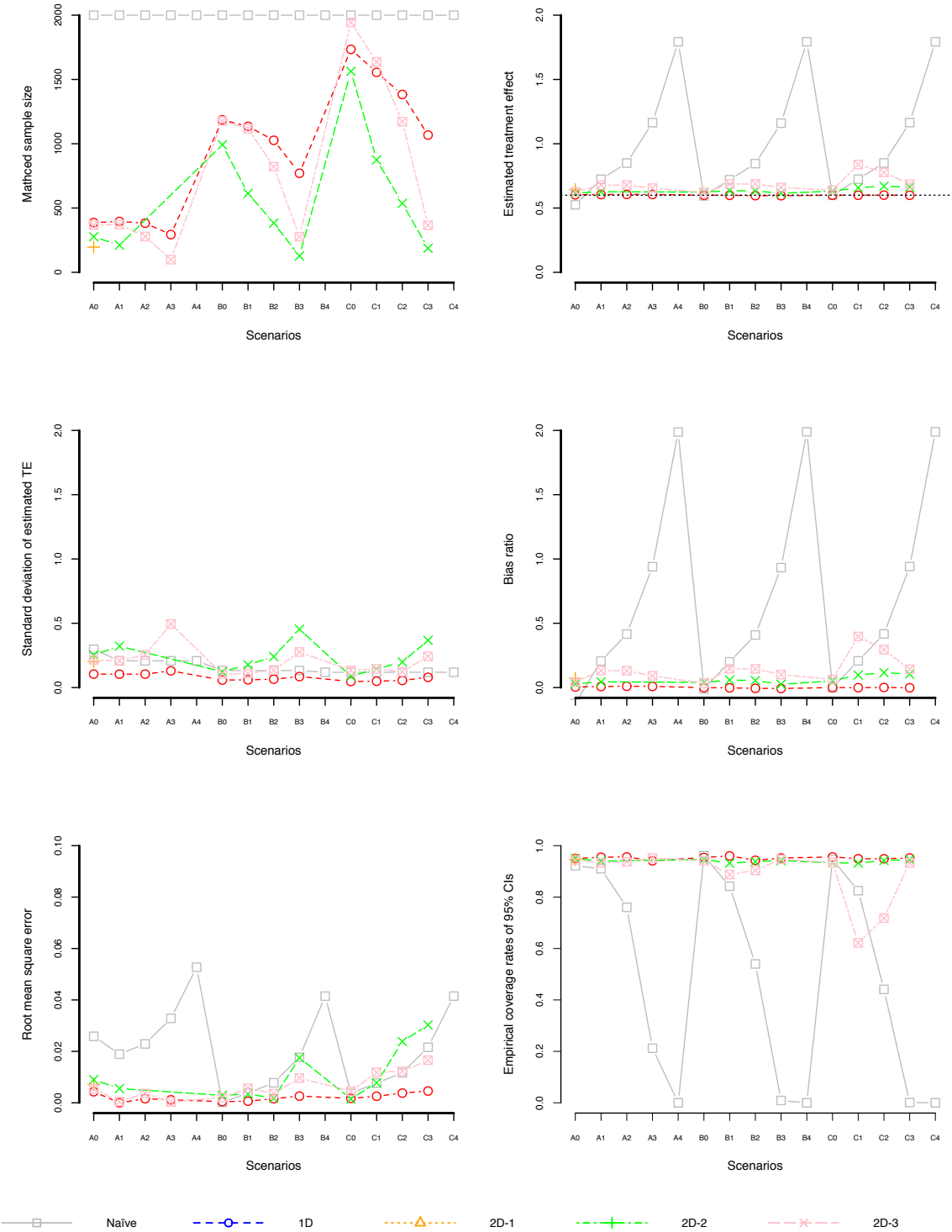

Figure 3 Performance measures of estimated average treatment effects

In a series of Monte Carlo simulations, we demonstrate 2DPSM is able to yield accurate and reliable estimates of the treatment effects in nonexperimental and non-panel settings in which the subjects in the treatment group and control group differ substantially, and the pool of subjects is different over time. The method is able to improve the causal inference by controlling *selection bias* and ensuring *longitudinal comparability*. For all matching methods, as the level of imbalance increases, the matched sample size decreases; as the prevalence of treatment increases, the matched sample size increases, but the level of balance after matching decreases. When analyzing observational data, statistical modeling could improve the causal inference by accommodating remained imbalance between the control group and the treatment group. In contrast with the econometric rule of the trade-off between variance and bias, the bias in fact increase when the estimated standard deviation decreases when the larger caliper is chosen, for example, 2D-3 chooses one standard deviation of propensity score as the pruning criteria. The decreasing standard deviation results from the larger matched sample size, but it is only meaningful if the desired confidence interval coverage remains. Our results indicate that an ideal specification of matching methods and caliper could achieve optimized bias and CI coverage. For example, 2D-2 optimizes the bias and CI coverage in all its application scenarios.

## 6. Discussion and Application Guidelines
Applying the 2DPSM requires researchers to make a few decisions. To make reasonable decisions, we need to understand the strength and weakness of different specifications, which is essentially trade-offs between estimation efficiency and bias. The implementation of 2DPSM should follow the conventional PSM practice but with more attention to the extra complexity. How might a researcher use the method in other settings? We provide our guidance for the implementation of the 2DPSM built upon Caliendo and Kopeinig (2008). Figure 4 presents a decision process for 2DPSM implementation.

The first and most important issue is whether the assumptions of propensity score matching method and difference-in-differences framework are valid. Namely, the assumptions are the ignorability that is the treatment selection process depends on variables that are observable (Rubin, 1977), and the common trend that is the treated group and control group have the same average treatment effects from the treatment (Athey and Imbens, 2006). 2DPSM comes to play only when the researcher is comfortable with the two assumptions.

Second, the overall quality of the data, particularly the quality of the control group, substantially influences the quality of the estimates. Essentially, matching is to find a comparable group for the treated group. In observational studies, the research would not be able to know the true treatment effect. Therefore, it is extremely valuable to check the comparability of the treated group and the control group (Dehejia and Wahba, 2002; Heckman et al., 1997a). In other words, it is critical to check whether there is sufficient overlap of the propensity score distribution in the region of common support to avoid comparing with the incomparable.

Then, one issue arises in implementing 2DPSM: which 2DPSM scheme to choose? We recommend 2D-1 as the first choice. 2D-1 conducts matching across both longitudinal and cross-sectional dimensions. This is beneficial in terms of self-selection bias reduction and comparability over time. Sometimes there may be few subjects remain during the matching process in this scheme when Mahalanobis distance matching algorithm used in longitudinal dimension, particularly when matching with rich information (i.e. many variables included). Under this condition, we recommend use 2D-2 or 2D-3.

If there is no sufficient overlap in the region of common support longitudinally, we recommend 1D scheme to avoid forcedly match subjects to subjects that are substantially different in terms of the estimated propensity score. This scheme can reduce the self-selection bias but requires parametric techniques to control for covariates. No matter which scheme used, imbalance checking after matching is important for researchers to get a better understanding of the extent to which the treatment and control groups comparable and thus of how sensitive estimates will be to the post-analysis model specifications.

Finally, researchers need to seek guidance on matching algorithm specifications such as caliper choice, whether or not match with replacement, number of control subjects to match to each treated subject. The general recommendations are 0.2 or 0.25 standard deviation of the estimated propensity score as caliper, no replacement, and no more than two control subjects to match to each treated subject. For detail guidelines, please refer to the work by Austin (2010, 2011b, 2014), Dehejia and Wahba (2002), and Caliendo and Kopeinig (2008).

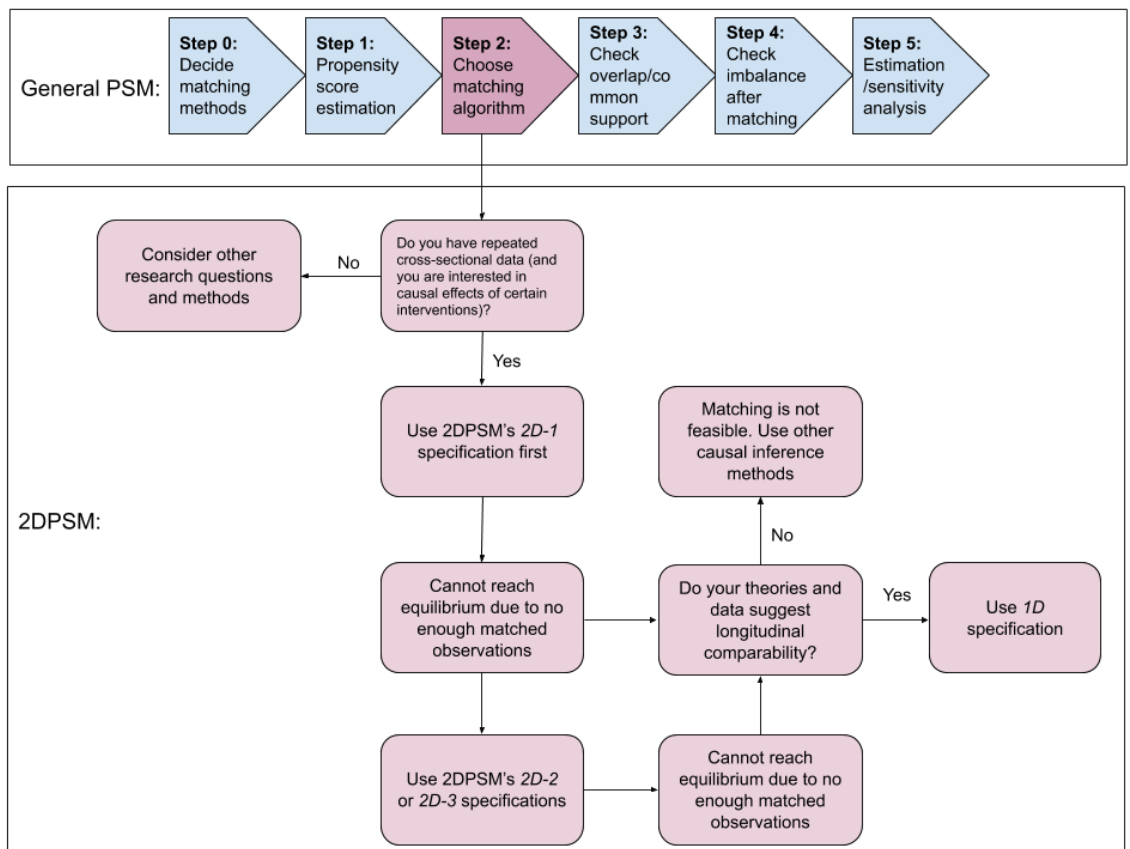

**Figure 4** Decision tree for 2DPSM implementation. The process of general PSM is modified from Figure 1 in Caliendo and Kopeinig (2008)

PSM is by no means a "magic bullet"(Smith and Todd, 2005), so does 2DPSM. In the 1D scheme, the correctly specified regression model yields highly accurate estimates. Under the ignorability assumption, propensity score matching is sufficient to summarize relevant information and address selection bias(Rosenbaum and Rubin, 1983). However, the true regression model is unknown in practice that strong parametric assumptions are needed for unobservable variables. Therefore, the performance of the 1D scheme is subject to the specification of statistical modeling when applied in observational studies. In addition, 2DPSM may not be applicable when the original sample size is small, and the level of imbalance is too high; in such situations, 2DPSM could be unable to be implemented, because the additional dimension of matching may prune more observations than one-dimensional matching and require larger common support of propensity score. The requirement for data quality is consistent with Heckman et al. (1998); Heckman et al. (1997a); Smith and Todd (2001) that matching methods perform reasonably well when applied to high-quality data. We note that matching at the longitudinal dimension is not necessary for scenarios with low imbalance levels that because our simulations satisfy the common trend assumption, and DID does not require true panel data if similar groups are followed over time (Blundell and Dias, 2009). However, populations around public transit stations would expect a dynamic change process due to the station opening and its induced development.

## 7. Conclusion

In this paper, we propose a new framework of applying matching methods at longitudinal and cross-sectional dimensions that establish pseudo-panel data from repeated cross-sections. It attempts to address data and causal inference barriers in the study of travel behavior and the built environment — specifically, 2DPSM pairs individuals from each of the before-after and treatment-control groups with similar characteristics.

This method is in the spirit of original matching methods in that it pairs identical individuals from the treated and control groups in order to ensure the balance of the covariates between those groups. It improves the original matching methods in two aspects. First, it offers a framework that reduces selection bias and improves longitudinal comparability. Second, it allows different matching algorithms to be implemented under this framework. Monte Carlo simulations demonstrate that the 2DPSM performs well in balancing data and improving estimates of the causal quantity of interest.

One caveat, which is also emphasized elsewhere(Dehejia and Wahba, 2002; Ho et al., 2007; Smith and Todd, 2001), is that matching should be viewed as a nonparametric preprocessing tool that complements the parametric analysis techniques. The 2DPSM can be extended to address heterogeneous, nonlinear, and dynamic treatment effects. Future research should build upon this framework to integrate new matching/classification algorithms that emerge in recent advances in statistics and machine learning literature.

## Acknowledgement

We received valuable comments and suggestions from various colleagues through the American Collegiate Schools of Planning (ACSP). This study is funded by the National Science Foundation (Award #: 1461766).

**Reference**

Abadie, A. (2005) 'Semiparametric difference-in-differences estimators', *The Review of Economic Studies*, 72(1), pp. 1-19

Althauser, R.P., Rubin, D. (1970) 'The computerized construction of a matched sample', *American Journal of Sociology*, pp. 325-346

Athey, S., Imbens, G.W. (2006) 'Identification and Inference in Nonlinear Difference-in-Differences Models', *Econometrica*, 74(2), pp. 431-497

Austin, P.C. (2010) 'Statistical criteria for selecting the optimal number of untreated subjects matched to each treated subject when using many-to-one matching on the propensity score', *American journal of epidemiology*, 172(9), pp. 1092-1097

Austin, P.C. (2011a) 'An introduction to propensity score methods for reducing the effects of confounding in observational studies', *Multivariate behavioral research*, 46(3), pp. 399-424

Austin, P.C. (2011b) 'Optimal caliper widths for propensity-score matching when estimating differences in means and differences in proportions in observational studies', *Pharmaceutical statistics*, 10(2), pp. 150-161

Austin, P.C. (2014) 'A comparison of 12 algorithms for matching on the propensity score', *Statistics in medicine*, 33(6), pp. 1057-1069

Blundell, R., Dias, M.C. (2009) 'Alternative approaches to evaluation in empirical microeconomics', *Journal of Human Resources*, 44(3), pp. 565-640

Boarnet, M.G. (2011) 'A broader context for land use and travel behavior, and a research agenda', *Journal of the American Planning Association*, 77(3), pp. 197-213

Caliendo, M., Kopeinig, S. (2008) 'Some practical guidance for the implementation of propensity score matching', *Journal of economic surveys*, 22(1), pp. 31-72

Cao, X., Fan, Y. (2012) 'Exploring the influences of density on travel behavior using propensity score matching', *Environment and Planning Part B*, 39(3), pp. 459

Cao, X., Schoner, J. (2014) 'The influence of light rail transit on transit use: An exploration of station area residents along the Hiawatha line in Minneapolis', *Transportation Research Part A: Policy and Practice*, 59(pp. 134-143

Cao, X., Xu, Z., Fan, Y. (2010) 'Exploring the connections among residential location, self-selection, and driving: Propensity score matching with multiple treatments', *Transportation Research Part A: Policy and Practice*, 44(10), pp. 797-805

Cao, X.J. (2010) 'Exploring causal effects of neighborhood type on walking behavior using stratification on the propensity score', *Environment and Planning A*, 42(2), pp. 487-504

Card, D. (1990) 'The impact of the Mariel boatlift on the Miami labor market', *ILR Review*, 43(2), pp. 245-257

Card, D., Krueger, A.B. (1993) 'Minimum wages and employment: A case study of the fast food industry in New Jersey and Pennsylvania', National Bureau of Economic Research.

Chatman, D.G. (2014) 'Estimating the effect of land use and transportation planning on travel patterns: Three problems in controlling for residential self-selection', *Journal of Transport and Land use*, 7(3), pp. 47-56

Cochran, W.G., Rubin, D.B. (1973) 'Controlling bias in observational studies: A review', *Sankhyā: The Indian Journal of Statistics, Series A*, pp. 417-446

d'Agostino, R.B. (1998) 'Tutorial in biostatistics: propensity score methods for bias reduction in the comparison of a treatment to a non-randomized control group', *Stat Med*, 17(19), pp. 2265-2281

Dargay, J.M., Vythoulkas, P.C. (1999) 'Estimation of a dynamic car ownership model: a pseudo-panel approach', *Journal of Transport Economics and Policy*, pp. 287-301

Deaton, A. (1985) 'Panel data from time series of cross-sections', *Journal of econometrics*, 30(1-2), pp. 109-126

Dehejia, R.H., Wahba, S. (2002) 'Propensity score-matching methods for nonexperimental causal studies', *Review of Economics and statistics*, 84(1), pp. 151-161

Garvey, G.T., Hanka, G. (1999) 'Capital structure and corporate control: The effect of antitakeover statutes on firm leverage', *The Journal of Finance*, 54(2), pp. 519-546

Gu, X.S., Rosenbaum, P.R. (1993) 'Comparison of multivariate matching methods: Structures, distances, and algorithms', *Journal of Computational and Graphical Statistics*, 2(4), pp. 405-420

Guo, S., Fraser, M.W. (2014) *Propensity score analysis: Statistical methods and applications*. Sage Publications.

Handy, S., Cao, X., Mokhtarian, P.L. (2006) 'Self-selection in the relationship between the built environment and walking: Empirical evidence from Northern California', *Journal of the American Planning Association*, 72(1), pp. 55-74

Heckman, J., Ichimura, H., Smith, J., Todd, P. (1998) 'Characterizing selection bias using experimental data', National bureau of economic research.

Heckman, J., Ichimura, H., Todd, P.E. (1997a) 'Matching as an econometric evaluation estimator: Evidence from evaluating a job training programme', *The review of economic studies*, 64(4), pp. 605-654

Heckman, J.J., Ichimura, H., Todd, P.E. (1997b) 'Matching as an econometric evaluation estimator: Evidence from evaluating a job training programme', *The review of economic studies*, 64(4), pp. 605-654

Heinze-Deml, C., Maathuis, M.H., Meinshausen, N. (2018) 'Causal structure learning', *Annual Review of Statistics and Its Application*, 5(pp. 371-391

Ho, D.E., Imai, K., King, G., Stuart, E.A. (2007) 'Matching as nonparametric preprocessing for reducing model dependence in parametric causal inference', *Political analysis*, 15(3), pp. 199-236

Holland, P.W. (1986) 'Statistics and causal inference', *Journal of the American statistical Association*, 81(396), pp. 945-960

Honey-Rosés, J., Stevens, M. (2017) 'Commentary on the Absence of Experiments in Planning', *Journal of Planning Education and Research*, pp. 0739456X17739352

Imai, K., King, G., Stuart, E.A. (2008) 'Misunderstandings between experimentalists and observationalists about causal inference', *Journal of the royal statistical society: series A (statistics in society)*, 171(2), pp. 481-502

Imbens, G., Liebman, J.B., Eissa, N. (1997) 'The Econometrics of Difference in Differences', *Harvard University Department of Economics. Mimeo*,

Imbens, G.W. (2004) 'Nonparametric estimation of average treatment effects under exogeneity: A review', *Review of Economics and statistics*, 86(1), pp. 4-29

King, G., Keohane, R.O., Verba, S. (1994) *Designing social inquiry: Scientific inference in qualitative research*. Princeton university press.

King, G., Nielsen, R. (2016) 'Why propensity scores should not be used for matching', *Copy at http://j. mp/1sexgVw Download Citation BibTex Tagged XML Download Paper*, 378(

Kurth, T., Walker, A.M., Glynn, R.J., Chan, K.A., Gaziano, J.M., Berger, K., Robins, J.M. (2006) 'Results of multivariable logistic regression, propensity matching, propensity adjustment, and propensity-based weighting under conditions of nonuniform effect', *American Journal of Epidemiology*, 163(3), pp. 262-270

Lee, B.K., Lessler, J., Stuart, E.A. (2010) 'Improving propensity score weighting using machine learning', *Statistics in medicine*, 29(3), pp. 337-346

Lewis, D.K. (1973) 'Counterfactuals, Library of Philosophy and Logic', Basil Blackwell, Oxford.

Mahalanobis, P.C. (1936) 'On the generalized distance in statistics', *Proceedings of the National Institute of Sciences (Calcutta)*, 2(pp. 49-55

McCormack, G.R., Shiell, A. (2011) 'In search of causality: a systematic review of the relationship between the built environment and physical activity among adults', *International Journal of Behavioral Nutrition and Physical Activity*, 8(1), pp. 1

Meyer, B.D. (1995) 'Natural and quasi-experiments in economics', *Journal of business & economic statistics*, 13(2), pp. 151-161

Meyer, B.D., Viscusi, W.K., Durbin, D.L. (1995) 'Workers' compensation and injury duration: evidence from a natural experiment', *The American economic review*, pp. 322-340

Mokhtarian, P.L., Cao, X. (2008) 'Examining the impacts of residential self-selection on travel behavior: A focus on methodologies', *Transportation Research Part B: Methodological*, 42(3), pp. 204-228

Oakes, J.M., Johnson, P.J. (2006) 'Propensity score matching for social epidemiology', *Methods in social epidemiology*, 1(pp. 370-393

Rosenbaum, P.R., Rubin, D.B. (1983) 'The central role of the propensity score in observational studies for causal effects', *Biometrika*, 70(1), pp. 41-55

Rosenbaum, P.R., Rubin, D.B. (1985) 'Constructing a control group using multivariate matched sampling methods that incorporate the propensity score', *The American Statistician*, 39(1), pp. 33-38

Roy, A.D. (1951) 'Some thoughts on the distribution of earnings', *Oxford economic papers*, 3(2), pp. 135-146

Rubin, D.B. (1974) 'Estimating causal effects of treatments in randomized and nonrandomized studies', *Journal of educational Psychology*, 66(5), pp. 688

Rubin, D.B. (1977) 'Assignment to Treatment Group on the Basis of a Covariate', *Journal of educational Statistics*, 2(1), pp. 1-26

Rubin, D.B. (1980) 'Bias reduction using Mahalanobis-metric matching', *Biometrics*, pp. 293-298

Rubin, D.B. (2005) 'Causal inference using potential outcomes: Design, modeling, decisions', *Journal of the American Statistical Association*, 100(469), pp. 322-331

Setoguchi, S., Schneeweiss, S., Brookhart, M.A., Glynn, R.J., Cook, E.F. (2008) 'Evaluating uses of data mining techniques in propensity score estimation: a simulation study', *Pharmacoepidemiology and drug safety*, 17(6), pp. 546-555

Smith, J.A., Todd, P.E. (2001) 'Reconciling conflicting evidence on the performance of propensity-score matching methods', *The American Economic Review*, 91(2), pp. 112-118

Smith, J.A., Todd, P.E. (2005) 'Does matching overcome LaLonde's critique of nonexperimental estimators?', *Journal of econometrics*, 125(1), pp. 305-353

Spears, S., Boarnet, M.G., Houston, D. (2016) 'Driving reduction after the introduction of light rail transit: Evidence from an experimental-control group evaluation of the Los Angeles Expo Line', *Urban Studies*, pp. 0042098016657261

Splawa-Neyman, J., Dabrowska, D.M., Speed, T. (1923) 'On the application of probability theory to agricultural experiments. Essay on principles. Section 9 (translated in 1990)', *Statistical Science*, 5(4), pp. 465-480

Stuart, E.A. (2010) 'Matching methods for causal inference: A review and a look forward', *Statistical science: a review journal of the Institute of Mathematical Statistics*, 25(1), pp. 1

Verbeek, M., Vella, F. (2005) 'Estimating dynamic models from repeated cross-sections', *Journal of econometrics*, 127(1), pp. 83-102

Xu, M., Yu, C.-Y., Lee, C., Frank, L.D. (2018) 'Single-Family Housing Value Resilience of Walkable Versus Unwalkable Neighborhoods During a Market Downturn: Causal Evidence and Policy Implications', *American Journal of Health Promotion*, 32(8), pp. 1714-1722

# Appendix A: Imbalance check after matching for each of the covariates

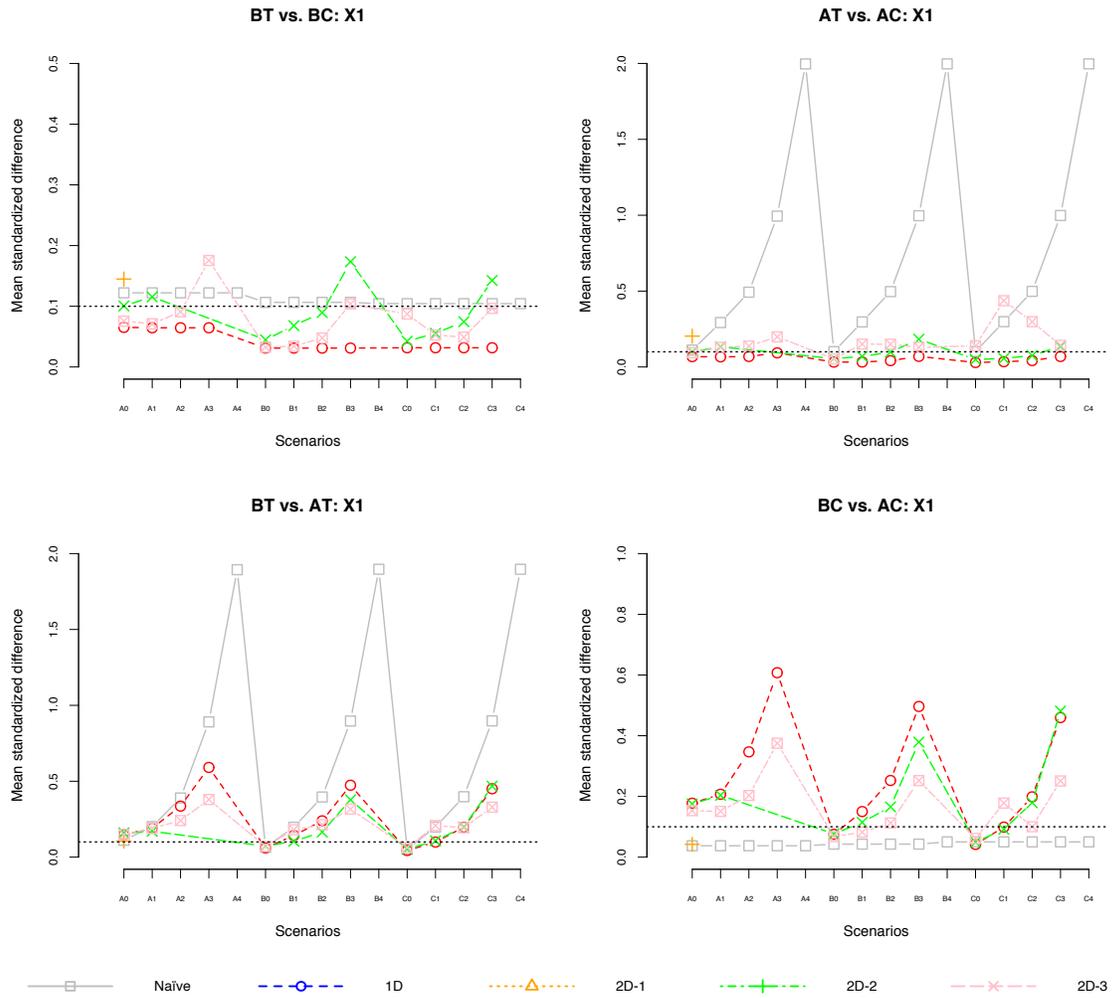

**Figure A1** Imbalance check of covariate X1.

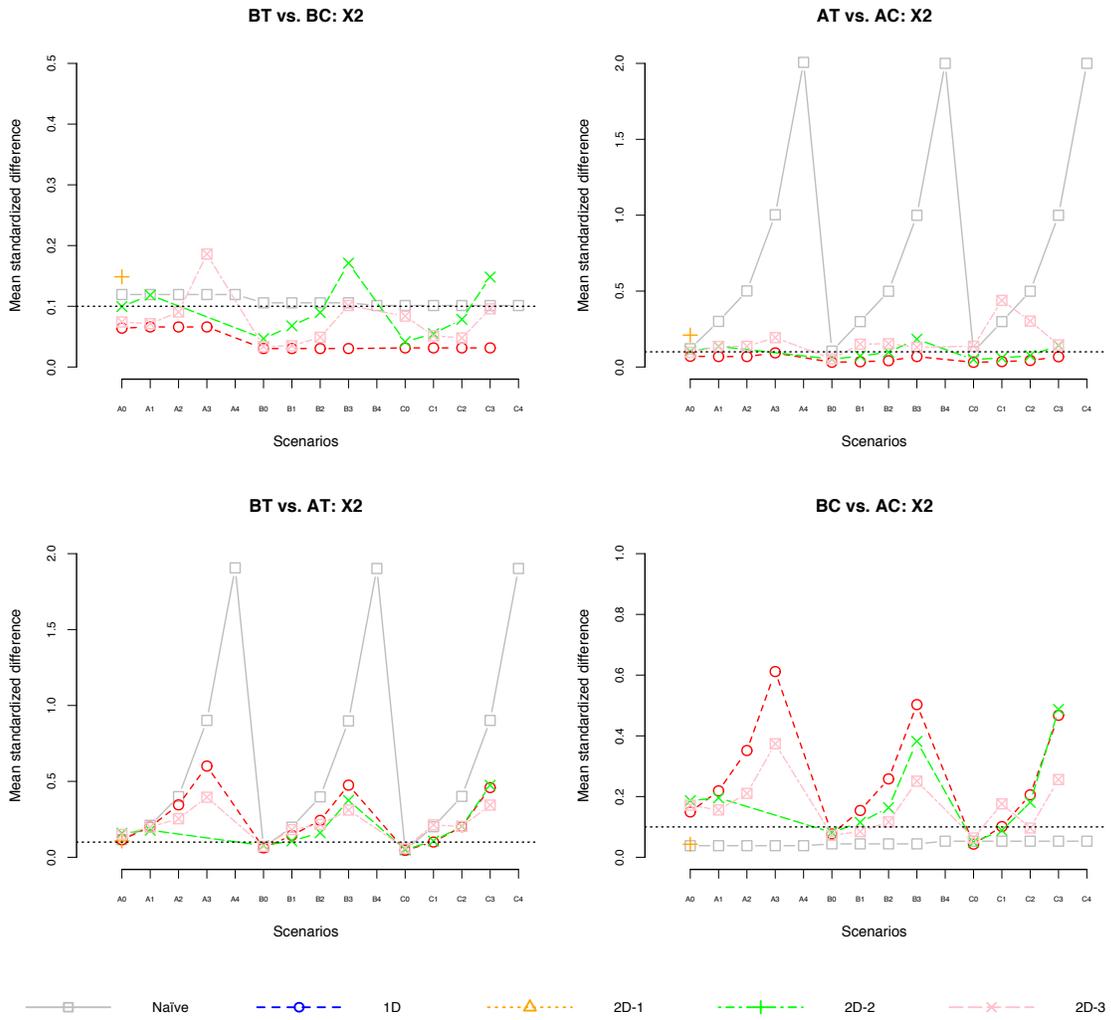

**Figure A2** Imbalance check of covariate X2.

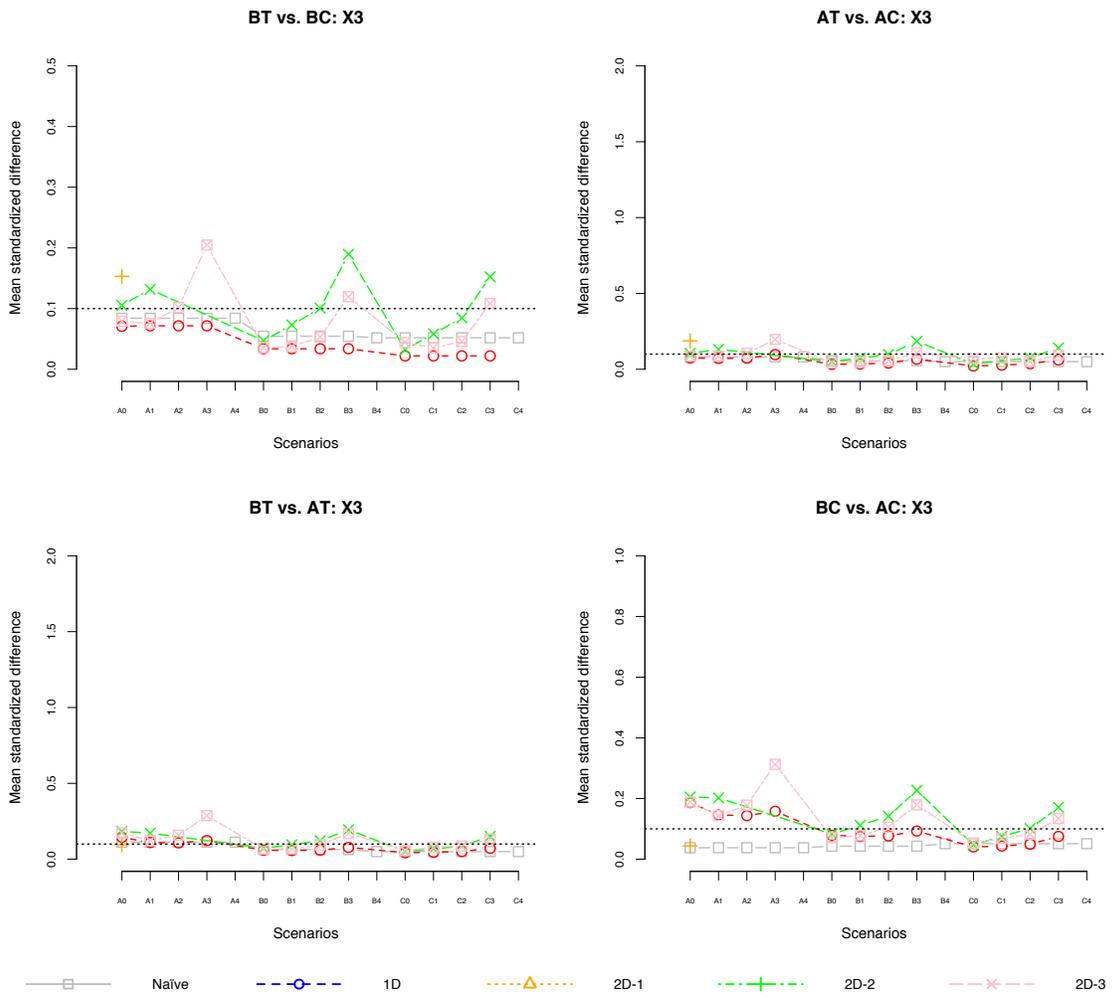

**Figure A3 I**mbalance of covariate X3.

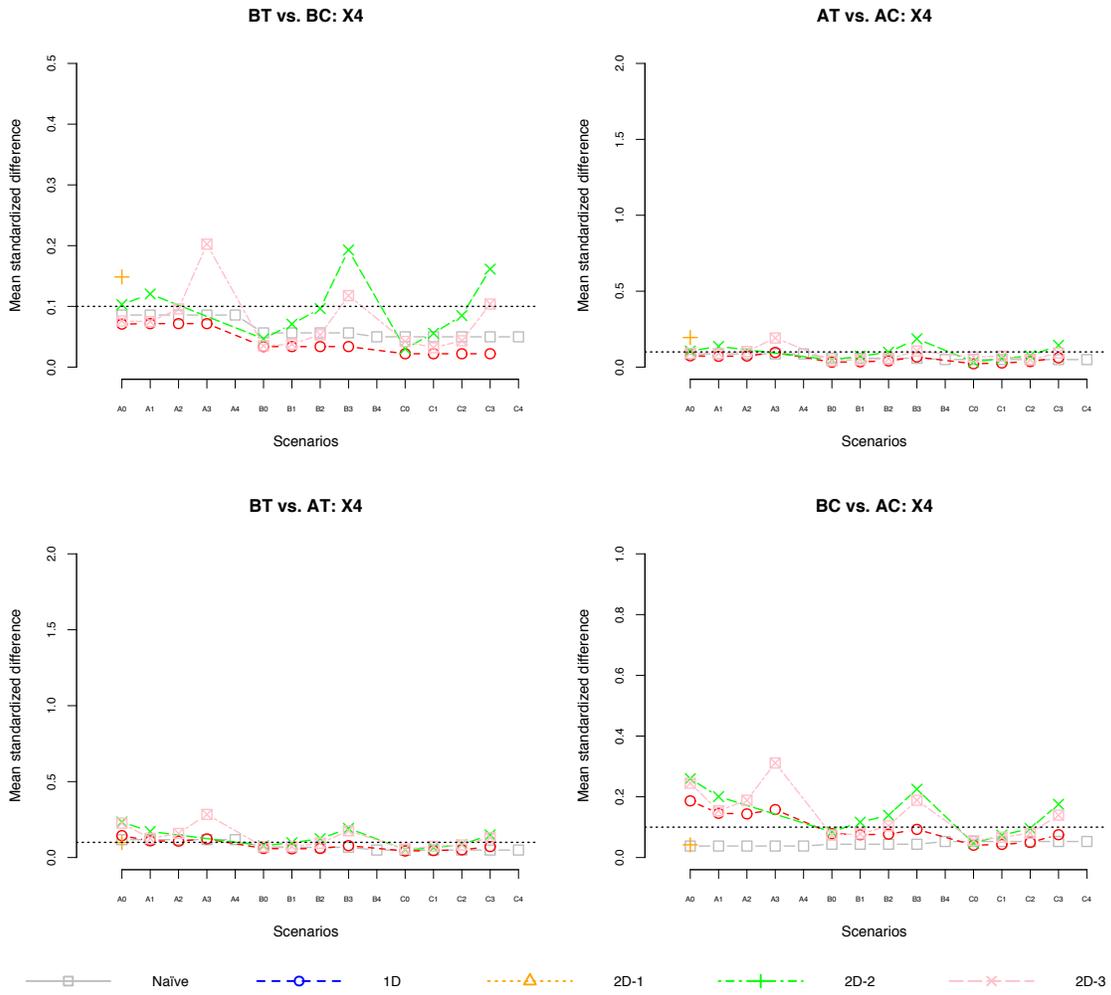

**Figure A4** Imbalance check of covariate X4.

**Appendix B: 2DPSM Performance Measures**

| Matching Scheme | Scenario | Matched Sample Size | Estimated Treatment Effect | SD of Estimated TE | Bias Ratio | RMSE | Coverage Rates of 95% CI |
|---|---|---|---|---|---|---|---|
| | A0 | 2000 | 0.53 | 0.30 | -0.12 | 0.03 | 0.92 |
| | A1 | 2000 | 0.72 | 0.21 | 0.21 | 0.02 | 0.91 |
| | A2 | 2000 | 0.85 | 0.21 | 0.42 | 0.02 | 0.76 |
| | A3 | 2000 | 1.16 | 0.21 | 0.94 | 0.03 | 0.21 |
| | A4 | 2000 | 1.79 | 0.21 | 1.99 | 0.05 | 0.00 |
| | B0 | 2000 | 0.59 | 0.13 | -0.01 | 0.00 | 0.96 |
| | B1 | 2000 | 0.72 | 0.13 | 0.20 | 0.00 | 0.84 |
| Naïve | B2 | 2000 | 0.85 | 0.13 | 0.41 | 0.01 | 0.54 |
| | B3 | 2000 | 1.16 | 0.13 | 0.93 | 0.02 | 0.01 |
| | B4 | 2000 | 1.79 | 0.12 | 1.99 | 0.04 | 0.00 |
| | C0 | 2000 | 0.60 | 0.12 | 0.00 | 0.00 | 0.95 |
| | C1 | 2000 | 0.72 | 0.12 | 0.21 | 0.01 | 0.82 |
| | C2 | 2000 | 0.85 | 0.12 | 0.42 | 0.01 | 0.44 |
| | C3 | 2000 | 1.16 | 0.12 | 0.94 | 0.02 | 0.00 |
| | C4 | 2000 | 1.79 | 0.12 | 1.99 | 0.04 | 0.00 |
| | A0 | 387 | 0.60 | 0.10 | 0.00 | 0.00 | 0.95 |
| | A1 | 396 | 0.60 | 0.10 | 0.01 | 0.00 | 0.95 |
| | A2 | 382 | 0.61 | 0.10 | 0.01 | 0.00 | 0.96 |
| | A3 | 294 | 0.61 | 0.13 | 0.01 | 0.00 | 0.94 |
| | B0 | 1186 | 0.60 | 0.06 | 0.00 | 0.00 | 0.95 |
| | B1 | 1135 | 0.60 | 0.06 | 0.00 | 0.00 | 0.96 |
| 1D | B2 | 1027 | 0.60 | 0.07 | -0.01 | 0.00 | 0.94 |
| | B3 | 770 | 0.60 | 0.09 | -0.01 | 0.00 | 0.95 |
| | C0 | 1734 | 0.60 | 0.05 | 0.00 | 0.00 | 0.96 |
| | C1 | 1555 | 0.60 | 0.05 | 0.00 | 0.00 | 0.95 |
| | C2 | 1383 | 0.60 | 0.06 | 0.00 | 0.00 | 0.95 |
| | C3 | 1067 | 0.60 | 0.08 | 0.00 | 0.00 | 0.95 |
| 2D-1 | A0 | 196 | 0.64 | 0.20 | 0.07 | 0.01 | 0.95 |
| | A0 | 277 | 0.62 | 0.25 | 0.03 | 0.01 | 0.95 |
| | A1 | 211 | 0.63 | 0.32 | 0.04 | 0.01 | 0.94 |
| | B0 | 992 | 0.62 | 0.12 | 0.04 | 0.00 | 0.95 |
| | B1 | 614 | 0.63 | 0.18 | 0.06 | 0.00 | 0.93 |
| 2D-2 | B2 | 383 | 0.63 | 0.24 | 0.05 | 0.00 | 0.94 |
| | B3 | 128 | 0.61 | 0.45 | 0.02 | 0.02 | 0.94 |
| | C0 | 1563 | 0.63 | 0.09 | 0.05 | 0.00 | 0.93 |
| | C1 | 875 | 0.66 | 0.15 | 0.10 | 0.01 | 0.93 |
| | C2 | 536 | 0.67 | 0.20 | 0.12 | 0.02 | 0.94 |
| | C3 | 187 | 0.66 | 0.37 | 0.10 | 0.03 | 0.95 |
| | A0 | 368 | 0.63 | 0.22 | 0.04 | 0.01 | 0.94 |
| 2D-3 | A1 | 372 | 0.68 | 0.21 | 0.13 | 0.00 | 0.93 |
| | A2 | 279 | 0.68 | 0.25 | 0.13 | 0.00 | 0.94 |
| | A3 | 99 | 0.65 | 0.49 | 0.09 | 0.00 | 0.95 |

| | | | | | | |
|---|---|---|---|---|---|---|
| B0 | 1177 | 0.62 | 0.11 | 0.03 | 0.00 | 0.94 |
| B1 | 1117 | 0.69 | 0.11 | 0.15 | 0.01 | 0.89 |
| B2 | 822 | 0.69 | 0.14 | 0.15 | 0.00 | 0.90 |
| B3 | 277 | 0.66 | 0.28 | 0.10 | 0.01 | 0.95 |
| C0 | 1943 | 0.64 | 0.13 | 0.06 | 0.00 | 0.93 |
| C1 | 1637 | 0.84 | 0.14 | 0.40 | 0.01 | 0.62 |
| C2 | 1172 | 0.78 | 0.12 | 0.29 | 0.01 | 0.72 |
| C3 | 366 | 0.69 | 0.24 | 0.14 | 0.02 | 0.93 |